# "Distributed Human Identity: AI-Enabled Multi-Existence Through Cognitive Replication and Robotic Embodiments"

A S M Touhidul Islam

Professor Dr. John Tookey

Faculty of Design and Creative Technologies,

Auckland University of Technology, Auckland, New Zealand

Email: asm_touhidul_islam@yahoo.com; asm.touhidul.islam@autuni.ac.nz

**Abstract**

Human presence has traditionally been constrained by the limits of physical embodiment, allowing individuals to exist in only one place at a time. This article introduces Multi-Existence Identity (MEI)—a socio-technical framework that replicates cognitive, behavioral, and emotional attributes into AI-enabled embodiments capable of acting across digital and physical contexts in parallel. MEI advances beyond digital twins, telepresence, and multipresence avatars by embedding *cognitive fidelity, affective resonance, and contextual responsiveness* into distributed agents that function not only for, but as, the original individual. The framework integrates personality modeling, cognitive simulation, and a synchronization layer to maintain identity coherence across three embodiment channels: digital avatars, robotic embodiments, and agentic software agents. Differentiating itself from simulated assistants, MEI positions replicated identity as a dynamic and culturally situated extension of selfhood, foregrounding tacit engagement and relational authenticity. Application domains span professional work, education, healthcare, governance, family life, and media, offering transformative potential for productivity, caregiving, leadership, and creativity. Yet these opportunities also surface profound challenges concerning authenticity, consent, legal accountability, privacy, and the psychological meaning of presence. The article proposes a phased empirical roadmap to operationalize MEI through personality modeling, synchronization testing, robotic embodiment trials, and ethical stress-testing. By conceptualizing MEI as both a technological and cultural construct, the study reframes debates on identity and presence in digitally augmented societies, highlighting opportunities for human–AI integration while underscoring the need for inclusive ethical governance.

# 1. Introduction

Human presence is inherently constrained by physical embodiment, limiting individuals to a single location at any given time. This constraint forces trade-offs between overlapping obligations—such as professional meetings, family engagements, and social interactions—often at the expense of personal relationships, organizational efficiency, and leadership continuity (Steuer, 1992; Wang, 2018). While digital technologies have enabled forms of remote interaction, they remain limited in replicating the full spectrum of human cognition and behavior. Advances in artificial intelligence (AI), particularly in digital twins, telepresence systems, and social robotics, offer partial solutions by extending presence across digital and physical environments. However, these systems often lack emotional nuance, contextual adaptability, and identity continuity (Bruynseels et al., 2018; Dautenhahn, 2007; Ringeval et al., 2025).

Recent literature on distributed cognition and embodied AI has emphasized the need to reconceptualize human–machine interaction as a socio-technical system. Distributed cognition theory posits that cognitive processes are shared across individuals, tools, and environments (Hutchins, 1995; Jacobsen et al., 2025). Embodied AI research further suggests that cognition is grounded in sensorimotor experience and social context, challenging disembodied models of intelligence (Barrett & Stout, 2024). Despite these insights, current frameworks remain fragmented and insufficiently integrated. AI-supported remote operations, for example, face challenges in maintaining situational awareness, team coordination, and cognitive alignment between human and machine agents (Jacobsen et al., 2025).

This study introduces the concept of MEI, a theoretical framework for replicating human identity into AI-enabled embodiments capable of acting in parallel across diverse contexts. MEI draws on personality modeling and decision heuristics to support autonomous embodiments. The article makes three core contributions and argues that MEI is not merely a technical innovation but a redefinition of human presence.

   A. Conceptual Contribution: Introduces the Multi-Existence Identity (MEI) framework, distinguishing replicated identity from simulated agents and reframing human presence in socio-technical terms.

   B. Technical Contribution: Proposes a tri-channel embodiment architecture (digital avatars, robotic embodiments, agentic software agents) with a synchronization layer for identity coherence, supported by personality modeling and cognitive simulation.

   C. Empirical Contribution: Outlines a phased roadmap for operationalizing MEI through personality model development, avatar deployment, synchronization testing, robotic trials, and ethical stress-testing.

## 2. Related Work and Theoretical Foundations

The concept of MEI intersects several established domains of research that provide both technical foundations and philosophical grounding.

### 2.1 Digital Twins and Identity Replication

Digital twins—virtual replicas of physical entities—have been extensively deployed in fields such as manufacturing, construction, and healthcare (Tao et al., 2019; Fuller et al., 2020). Human-centered variants, such as patient twins for predictive healthcare (Bruynseels et al., 2018), highlight the feasibility of modeling behaviors and preferences in real time. Yet, these approaches remain limited by their reductionist orientation: individuals are rendered as datasets optimized for functional outcomes rather than as socially situated, emotionally responsive beings.

MEI extends beyond this limitation by shifting the focus from mechanical replication to identity continuity. Instead of treating the individual as a static model, MEI emphasizes dynamic personality modeling and replicated cognition, enabling a distributed but coherent replication of human identity. In doing so, MEI bridges the technical sophistication of digital twins with the richer social and philosophical dimensions of personhood.

### 2.2 Telepresence and Multipresence

Telepresence research has explored how individuals extend their sense of presence into remote environments (Steuer, 1992; Wang, 2018). Related work on "multipresence" has examined simultaneous presence across digital platforms (Haans & Ijssel Steijn, 2012). While valuable, these systems reduce presence to audiovisual continuity, offering representation without autonomy or authentic engagement. Scholars have further questioned whether telepresence can replicate the *felt* quality of co-presence, raising doubts about relational authenticity.

MEI addresses these deficiencies by integrating cognitive simulation and decision-making heuristics into its embodiments. Rather than transmitting a passive signal, MEI copies act with personality-consistent autonomy, allowing others to experience interaction not merely with a proxy but with a cognitively and emotionally resonant extension of the self.

2.2.1 Limitations of Existing Models

Despite their promise, existing models such as digital twins and telepresence systems face notable limitations when applied to human identity. Digital twins, while effective in modeling mechanical systems, are often criticized for their reductionist approach to human representation—treating individuals as data sets rather than complex, emotionally nuanced agents. This limits their capacity to capture the richness of human cognition and behavior. Similarly, telepresence technologies rely heavily on audiovisual transmission and lack the depth required for authentic social engagement. Scholars have questioned whether telepresence can ever replicate the *felt experience* of being with someone, raising concerns about agency, empathy, and relational authenticity. These critiques underscore the need for a more holistic framework—such as MEI—that integrates cognitive alignment, empathetic mirroring, and embodied interaction.

**2.3 Embodied AI and Social Robotics**

Social robotics and embodied AI research have demonstrated the feasibility of machines engaging with humans in naturalistic, socially meaningful ways (Breazeal, 2003; Dautenhahn, 2007). Robots equipped with conversational AI and affective computing capabilities can simulate empathy, maintain dialogues, and respond to environmental cues. Integrating personality modeling into such systems advances the potential for robotic embodiments that act as authentic extensions of an individual's identity.

**2.4 Philosophy of Identity and Distributed Cognition**

From a philosophical perspective, theories of identity emphasize continuity of consciousness and the persistence of memory as central to the notion of self (Locke, 1690/1975). Distributed cognition theories extend the human mind beyond the skull, embedding it within tools, systems, and social structures (Hutchins, 1995). MEI can thus be conceptualized as a technologically mediated extension of identity, where distributed agents embody fragments of cognition and agency across multiple contexts.

Importantly, MEI can also be understood through the lens of tacit engagement, where identity is not solely encoded as data but enacted through embodied, relational, and situational interactions. By modeling subtle emotional cues, behavioral rhythms, and unarticulated decision heuristics, MEI agents extend beyond explicit representation into the tacit dimensions of human presence. This positions MEI not only as a technical construct but also as a form of embodied cultural practice. Closely related constructs include the human digital twin (Lin et al., 2024), multipresence avatars (Haans & Ijssel Steijn, 2012), and the extended self in HCI (Turkle, 2011). While each addresses facets of distributed presence, they often remain either reductionist (digital twins), representational rather than autonomous (multipresence), or primarily psychological (extended self). MEI contributes by integrating these threads into a socio-technical framework that emphasizes implicit interaction and relational authenticity, extending debates into cultural and embodied domains. Taken together, the literatures on digital twins, telepresence, and social robotics illustrate partial pathways to distributed presence, but none achieve the integration of thought-style alignment, emotional resonance, and behavioral continuity. MEI is proposed as a framework that unifies these strands, moving the discourse from fragmented functionality to a coherent model of distributed human identity.

To further ground MEI in interdisciplinary discourse, this study draws on foundational work in tacit knowledge (Polanyi, 1966), embodied cognition (Varela et al., 1991; Sheets-Johnstone, 2011), and AI ethics (Floridi, 2018; Bryson, 2019; Bostrom, 2014). These literatures emphasize the importance of non-verbal, culturally embedded, and ethically governed interactions—dimensions central to MEI's design. Additionally, critiques of digital identity and surveillance (Zuboff, 2019; Lyon, 2007) highlight the risks of commodifying presence, reinforcing the need for MEI to prioritize consent, transparency, and relational authenticity.

**2.5 MEI's Differentiation From Existing Conceptual Models**

Several closely related constructs offer partial visions of distributed human presence yet fall short of MEI's holistic integration of cognitive fidelity, emotional resonance, and agentic autonomy. Distinction of MEI from Existing Concepts are presented in the Table 1.

- Human Digital Twin (HDT):

HDT frameworks (Lin et al., 2024; Song, 2023) typically model individuals as data-driven entities optimized for predictive analytics. While useful in healthcare and design, HDTs often reduce identity to quantifiable traits, lacking emotional nuance and situational adaptability. MEI responds by embedding non-verbal presence—subtle emotional rhythms, decision hesitations, and relational cues—into its embodiments, thereby resisting reductionist modeling.

- Multipresence Avatars:

Research on multipresence (Haans & Ijssel Steijn, 2012) explores simultaneous presence across platforms but remains largely representational. These avatars transmit audiovisual signals without autonomous decision-making or personality alignment. MEI advances this by enabling agentic autonomy—copies that act with consistent values and cognitive style, not merely as proxies.

- Extended Self in HCI:

Turkle (2011) conceptualizes the extended self as a psychological phenomenon shaped by digital interaction. However, MEI extends this into embodied and relational domains, emphasizing cultural embeddedness and situational responsiveness. MEI agents do not merely reflect internal states but enact identity through social and environmental engagement.

- Posthuman Identity in STS:

Drawing on posthumanist thinkers such as Hayles and Haraway, MEI can be framed as a techno-social assemblage that challenges anthropocentric boundaries. It reconfigures agency and embodiment by distributing selfhood across robotic, digital, and software agents. This positions MEI within broader debates on hybridity, autonomy, and the ethics of techno-human integration.

Table 1: Distinction of MEI from Existing Conceptual Models

| Concept | Core Limitation | MEI's Contribution |
|---|---|---|
| Human Digital Twin | Data-centric, lacks emotional and relational fidelity | Embeds tacit engagement and personality modeling |
| Multipresence Avatars | Representational, no autonomy or cognitive alignment | Enables agentic autonomy and decision-making consistency |
| Extended Self (HCI) | Psychological focus, lacks embodiment and cultural depth | Adds embodied, relational, and culturally situated presence |
| Posthuman Identity (STS) | Theoretical, lacks operational framework | Operationalizes techno-human hybridity via MEI embodiments |

## 2.6 Relation to Prior Work

The concept of Multi-Existence Identity (MEI) intersects with several established areas of research in autonomous agents and multi-agent systems, particularly in identity modeling, personality-driven behavior, synchronization, and embodied AI. While MEI introduces a novel socio-technical framework, it builds upon and extends foundational work in these domains.

- Identity and Personality Modeling in Agents

Efforts to embed personality traits and cognitive styles into autonomous agents have gained traction in recent years. Arukgoda et al. (2023) explored how agents endowed with motivational profiles—such as power, achievement, and affiliation—can exhibit diverse, risk-aware behaviors in collaborative environments. Their work demonstrates that personality-driven agents can enhance team performance and predictability, especially in high-risk domains. MEI builds on this by proposing a more granular personality modeling approach that includes emotional rhythms, decision heuristics, and cultural embeddedness, enabling agents to act not just predictably but authentically.

Gero et al. (2021) examined how users form mental models of AI agents in cooperative tasks, emphasizing the importance of behavioral consistency and conceptual transparency. MEI responds to this by integrating cognitive simulation and synchronization mechanisms that maintain continuity across multiple embodiments, thereby supporting richer and more intuitive human–AI interactions.

- Synchronization and Identity Coherence

Synchronization across distributed agents is a longstanding challenge in multi-agent systems. Saberi et al. (2022) provided a comprehensive treatment of synchronization in heterogeneous multi-agent systems, addressing issues such as delays, disturbances, and time-varying networks. MEI extends this technical foundation by introducing a bi-directional synchronization layer that not only aligns task execution but also preserves identity coherence—ensuring that distributed embodiments remain faithful to the original individual's intentions and values.

- Embodied AI and Distributed Cognition

Embodied multi-agent systems (EMAS) have emerged as a promising approach to handling complex, real-world tasks. Wu et al. (2025) conducted a systematic review of generative multi-agent collaboration in embodied AI, highlighting how foundation models can enhance perception, planning, and communication across physical and virtual agents. MEI contributes to this discourse by proposing a tri-channel embodiment architecture—digital avatars, robotic embodiments, and agentic software agents—each designed to enact identity through situated, relational engagement.

Unlike traditional EMAS frameworks that focus on task efficiency, MEI emphasizes relational authenticity and implicit interaction, drawing from distributed cognition theory to conceptualize identity as enacted across tools, environments, and social contexts.

## 3. Conceptual Framework: Multi-Existence Identity (MEI)

The proposed framework for MEI envisions a system in which an individual's identity is captured, modeled, and deployed into multiple autonomous embodiments that operate in parallel.

### 3.1 Core Components of MEI

A. Personality Modeling: MEI requires a core model trained on linguistic patterns, heuristics, and emotional responses. Unlike digital assistants, which operate on generic datasets, MEI's personality model encodes the individual's value system and interaction style, ensuring that distributed embodiments act not only *for* but *as* the person.

B. Cognitive Simulation: To achieve behavioral fidelity, MEI incorporates algorithms capable of simulating decision latencies, hesitations, and non-verbal cues. These micro-dynamics are essential for authentic social interaction and prevent MEI from collapsing into mere scripted responses.

C. Embodiment Channels: MEI manifests through three interconnected channels:

   o *Digital Avatars* for online platforms and metaverse settings.

- o *Robotic Embodiments* providing tangible, physical representation.

- o *Agentic Software Agents* performing parallel cognitive or communicative tasks. Each channel operationalizes a different dimension of identity, making MEI more robust than any single modality.

D. Synchronization Layer: Coherence across embodiments requires a bi-directional synchronization system. Without such a mechanism, copies risk divergence from the individual's core intentions. Synchronization ensures continuity of memory, identity alignment, and override capacity. This design responds directly to the fragmentation issues seen in telepresence and avatar-based systems.

E. Conceptual Boundary: Replicated Identity vs. Simulated Agent

To clarify the conceptual scope of MEI, it is important to distinguish between replicated identity and simulated agents. A replicated identity refers to an MEI agent that mirrors the original individual's cognitive style, emotional responses, and decision-making logic with high fidelity. Such agents are designed to act *as* the individual, maintaining continuity of personality and values across contexts. In contrast, a simulated agent is a generic AI assistant or avatar that performs tasks based on programmed instructions or generalized behavior, without deep personalization or identity continuity. MEI aspires to transcend simulation by achieving cognitive and behavioral fidelity, thereby functioning as a distributed extension of selfhood rather than a mere tool.

**3.2 Functional Logic of MEI**

- Task selection and autonomy:

MEI deployment begins with task selection, where the individual defines which activities require direct presence and which may be delegated. Replicated embodiments operate with bounded autonomy, constrained by personality parameters and synchronization protocols. This logic echoes debates in distributed agency and AI governance, where autonomy is balanced by accountability

- Feedback and endorsement:

  - o Embodied agents can relay summaries back to the original individual, integrating distributed experiences into a coherent memory stream.

  - o Delegation requires mechanisms for feedback and endorsement to ensure the copy's actions align with the source identity.

  - o Endorsement protocols allow actions to be reviewed, ratified, or overridden by the individual, maintaining coherence, accountability, and trust.

- Escalation for critical situations:
    - MEI embodiments must incorporate escalation protocols for unforeseen or high-stakes scenarios.
    - Copies can consult the original individual in real-time without disrupting outward continuity of interaction.
    - This "silent escalation" balances autonomy with authenticity, preventing overstepping while preserving social immersion.
- Organizational deployment and authority:
    - Deploying MEI in hierarchical organizations can create role conflicts and ambiguity in authority recognition.
    - For example, a delegated MEI agent chairing a meeting may be perceived as undermining deputies or second-in-command officers.
    - Explicit protocols are needed to define when MEI may act in place of a leader, ensuring deputies' roles are respected and organizational norms are maintained.

### 3.3 Differentiation from Existing Technical Models

Unlike digital assistants or simple avatars, MEI aspires to integrate mental coherence (thinking like the original), behavioral fidelity (acting like the original), and contextual adaptability (making decisions aligned with the original's values). This transforms AI from a support tool into a *distributed extension of selfhood*.

Table 2: Distinction of MEI from Existing Technical Models

| Existing Approach | Limitation | How MEI Responds |
|---|---|---|
| Digital Twins | Reductionist, data-driven; little emotional/social nuance | Encodes personality, cognition, and emotional fidelity |
| Telepresence | Audiovisual continuity only; lacks autonomy and authenticity | Provides autonomous, personality-consistent decision-making |
| Social Robotics | Empathy simulation without identity continuity | Embeds original's personality model into robotic form |

| Existing Approach | Limitation | How MEI Responds |
|---|---|---|
| Generic AI Assistants | Task automation, no identity alignment | Functions as a distributed extension of selfhood |

## 3.4 Proposed Visual Conceptual Model for MEI

Figure 1 (below) presents a compact conceptual model for the Multi-Existence Identity (MEI) framework. At the centre is the Original Identity (the physical human self). Immediately surrounding the centre is the AI Personality Model — a dynamic model that encodes the individual's cognitive heuristics, affective patterns, and decision heuristics. From the personality model three embodiment channels extend outward and operate in parallel:

- Digital Avatars — virtual representations used in online platforms, virtual meetings, and metaverse environments.

- Robotic Embodiments — physical robots (humanoid or service robots) that interact in real world settings.

- Agentic Software Agents — autonomous programs that perform tasks (negotiation, scheduling, companionship) on behalf of the person.

A Synchronization Layer maintains bi-directional memory, state and decision alignment between the Original Identity and all embodiments, enforcing coherence and providing override/endorsement mechanisms. Surrounding these components is a band of Application Domains (professional work, family/social life, education, healthcare, governance), indicating where MEI embodiments may be deployed.

The diagram shows: Original Identity (center) → AI Personality Model → three Embodiment Channels (Digital Avatars, Robotic Embodiments, Agentic Software Agents) with a Synchronization Layer linking them back to the Original Identity; Application Domains are shown as the contextual ring around the system.

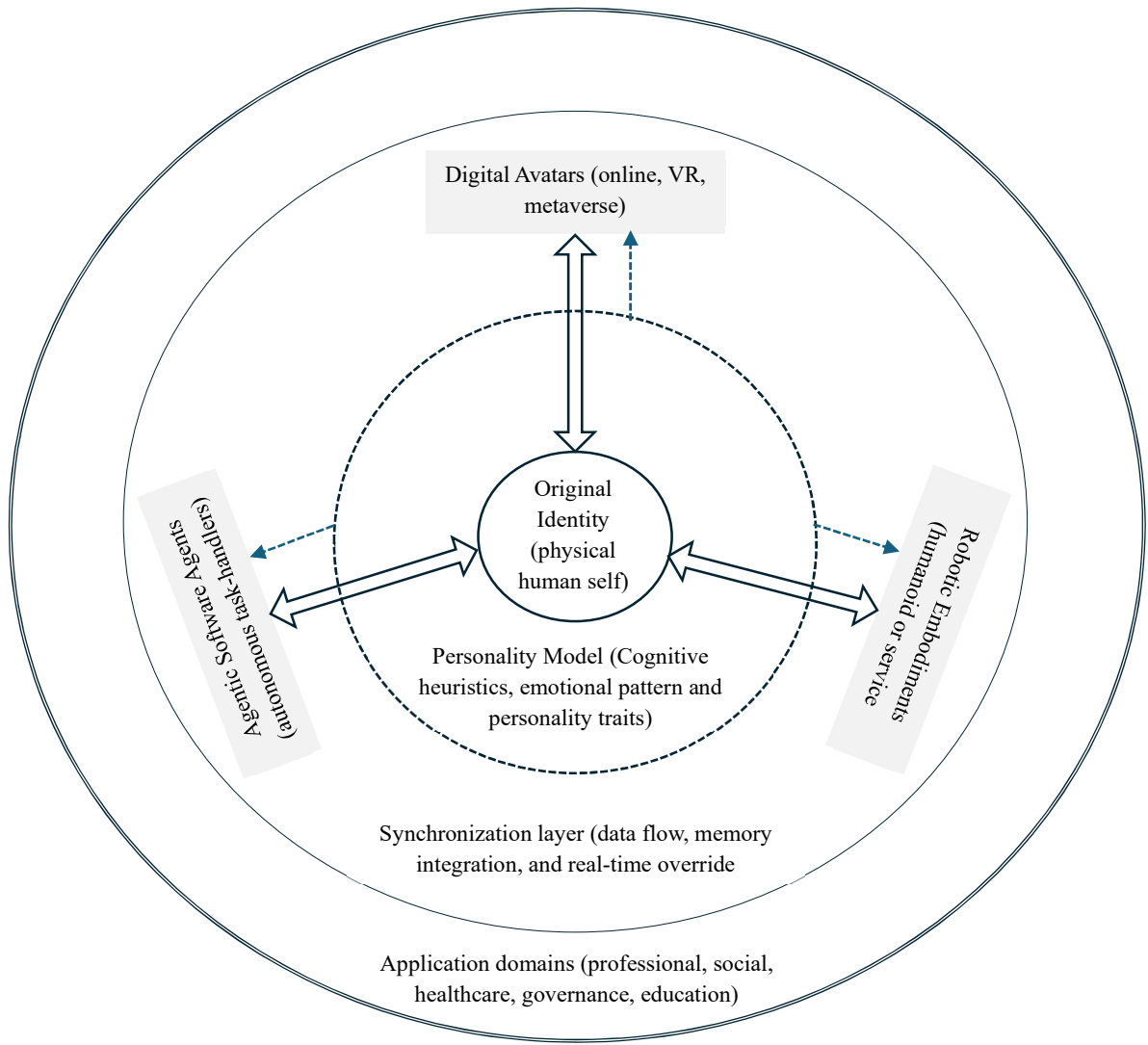

**Figure 1.** Conceptual Model of MEI (Author's creation)

### 3.5 Technical Contributions of MEI

The MEI framework contributes to the field of autonomous agents and multi-agent systems through several novel technical elements that distinguish it from existing models:

### A. Identity-Centric Personality Modeling

Unlike generic AI assistants, MEI agents are built on a person-specific personality model trained using linguistic patterns, emotional responses, and decision heuristics. This model enables agents to act with cognitive fidelity—thinking and responding in ways consistent with the original individual's values and style.

- Technical novelty: Use of fine-tuned NLP models (e.g., GPT, BERT) on individual corpora to simulate personality traits.

- Contribution: Moves beyond task-based automation to identity-based autonomy.

### B. Multi-Channel Embodiment Architecture

MEI introduces a tri-channel embodiment system:

- Digital avatars for virtual platforms.

- Robotic embodiments for physical interaction.

- Agentic software agents for autonomous task execution.

Each channel is synchronized to maintain identity coherence, allowing distributed agents to operate in parallel without divergence.

- Technical novelty: Integration of embodiment modalities with a shared personality core.

- Contribution: Enables parallel presence with consistent behavior across contexts.

### C. Synchronization Layer for Identity Coherence

A key innovation is the bi-directional synchronization layer that ensures memory, emotional state, and decision logic remain aligned across all MEI embodiments.

- Technical novelty: Real-time synchronization protocols inspired by distributed cognition and multi-agent reinforcement learning.

- Contribution: Prevents fragmentation and drift in agent behavior, supporting long-term coherence.

### D. Functional Logic for Delegation and Escalation

MEI agents operate with bounded autonomy, guided by endorsement and escalation protocols:

- Endorsement: Agents report back and seek approval for critical decisions.

- Escalation: In high-stakes scenarios, agents silently consult the original individual.

- Technical novelty: Embedded feedback loops and override mechanisms.

- Contribution: Balances autonomy with accountability, a key challenge in MAS governance.

### E. Ethical Stress-Testing Framework

MEI includes a roadmap for ethical and legal stress-testing, simulating real-world scenarios such as contract signing, emotional interaction, and data governance.

- Technical novelty: Use of mock legal trials, Delphi panels, and GDPR-aligned protocols.
- Contribution: Embeds ethics into the design and deployment lifecycle of autonomous agents.

**3.6 Mapping MEI to Technical Paradigms in Multi-Agent Systems**

The Multi-Existence Identity (MEI) framework shares conceptual and operational affinities with several foundational paradigms in multi-agent systems (MAS). These include agent-based modeling, reinforcement learning, belief–desire–intention (BDI) architectures, and multi-agent coordination. By situating MEI within these established technical domains, the framework can be understood not only as a socio-technical construct but also as a contribution to the design of cognitively and emotionally coherent autonomous agents.

A. **Agent-Based Modeling (ABM)**

MEI agents are instantiated as autonomous entities with distinct behavioral rules derived from personality models. This mirrors ABM, where agents operate based on internal states and interaction rules within a simulated environment (Wooldridge, 2009). MEI extends ABM by embedding identity fidelity and cognitive simulation, enabling agents to act not just as functional units but as distributed extensions of human identity.

B. **Reinforcement Learning (RL)**

MEI agents can use RL to adapt their behavior based on feedback from interactions. For example, an MEI agent deployed in a customer service role could learn optimal responses while preserving personality traits. Synchronization mechanisms in MEI can be tested using multi-agent RL to ensure coherence across embodiments (Busoniu et al., 2008). MEI introduces bounded autonomy, where RL is constrained by personality parameters and override protocols, balancing learning with identity preservation.

C. **Belief–Desire–Intention (BDI) Models**

BDI models represent agents with internal mental states—beliefs, desires, and intentions—that guide decision-making (Rao & Georgeff, 1995). MEI agents similarly operate based on encoded values, emotional rhythms, and situational goals. MEI enriches BDI with affective resonance and cultural embeddedness, making agent decisions socially and emotionally authentic.

D. **Multi-Agent Coordination**

MEI agents deployed across different contexts (e.g., healthcare, education, governance) must coordinate actions to maintain identity coherence. This requires protocols for task delegation, feedback integration, and escalation. MEI adds a synchronization layer that ensures not just task alignment but identity alignment, a novel contribution to MAS coordination (Durfee, 1999).

**Table 3:** MEI and MAS Paradigms

| MAS Paradigm | MEI Alignment | MEI Extension |
|---|---|---|
| Agent-Based Modeling | Autonomous agents with personality-driven rules | Adds cognitive fidelity and identity replication |
| Reinforcement Learning | Adaptive behavior based on feedback | Bounded by personality and override protocols |
| BDI Models | Belief–desire–intention logic for decision-making | Enriched with emotional and cultural dimensions |
| Multi-Agent Coordination | Distributed agents maintaining coherence | Synchronization of identity and memory streams |

## 4. Applications of Multi-Existence Identity (MEI)

The implementation of MEI presents transformative potential across multiple domains by enabling distributed presence and cognitive replication. This section explores six key application areas, each supported by recent scholarly literature.

### 4.1 Professional and Organizational Contexts

In professional environments, MEI offers a paradigm shift in how individuals manage concurrent responsibilities. In organizational contexts, MEI embodiments could theoretically extend presence across multiple engagements, raising questions about continuity, visibility, and authority structures. This reframing aligns with debates in organizational AI governance (Acosta & Hughes, 2025). Recent studies emphasize the strategic integration of AI agents in corporate decision-making and workflow optimization, underscoring the value of MEI in augmenting human capital (Acosta & Hughes, 2025; Boch & Thomas, 2025).

Moreover, MEI can support adaptive governance frameworks, allowing leaders to synthesize policy options, monitor implementation, and engage stakeholders simultaneously. RAND Corporation's model of AI-enabled governance illustrates how distributed agents can enhance administrative efficiency and participatory decision-making (Acosta & Hughes, 2025). MEI embodiments, when aligned with organizational values and ethical protocols, can serve as trusted extensions of leadership, facilitating agile responses to complex challenges.

## 4.2 Family and Social Life

In family settings, emotional continuity may be preserved through MEI agents that replicate familiar tones and gestures and companionship within family systems, particularly during periods of physical absence. Robotic embodiments infused with personalized traits may engage with children, elderly relatives, or partners, promoting cohesion and relational stability. In family life, MEI illustrates tensions between affective continuity and authenticity. For example, agents modeled on parental interaction styles highlight how AI-mediated intimacy can sustain relational bonds while simultaneously raising ethical concerns (Bozdağ, 2025). This emotionally faithful interaction underscores how MEI agents can preserve relational authenticity by replicating not just words, but the emotional alignment embedded in family life. In this way, emotional fidelity becomes a core dimension of MEI, strengthening bonds despite physical absence. Yang & Oshio (2025) demonstrate that AI applications can enhance family communication and support value-based interactions, while Bozdağ (2025) highlights the role of personalization and accessibility in strengthening familial bonds.

However, relational embodiments—such as simulating a former partner—require ethical scrutiny. While such applications may offer comfort or therapeutic value, they risk emotional dependency and identity distortion. Scholars caution against overreliance on AI-mediated intimacy, emphasizing the need for transparency, consent, and psychological safeguards (Bozdağ, 2025; Yang & Oshio, 2025). MEI systems must be designed to respect emotional boundaries and promote healthy social dynamics.

## 4.3 Education and Training

In educational contexts, MEI agents can participate in parallel learning environments, enabling students to engage with multiple courses or workshops simultaneously. AI-powered embodiments may record sessions, interact with instructors, and relay insights to the original individual. Jose et al. (2025) explore the cognitive paradox of AI in education, noting its potential to enhance personalization while cautioning against cognitive offloading and reduced engagement.

Bauer et al. (2025) propose the ISAR model, which categorizes AI's impact on learning as inversion, substitution, augmentation, or redefinition. MEI agents, when designed to augment rather than replace cognitive effort, can support higher-order thinking and critical analysis. Educators may also deploy MEI embodiments to extend their reach across classrooms, promoting equitable access and pedagogical innovation.

## 4.4 Healthcare and Well-Being

MEI offers significant promise in healthcare by enabling distributed consultations, personalized monitoring, and psychosocial support. AI-integrated digital twins can simulate patient conditions, optimize treatment plans, and facilitate real-time interventions. Basu (2025) and Bruynseels et al. (2018)

underscore the role of AI-driven digital twins in precision medicine, highlighting their capacity to model health trajectories and personalize care.

Patient-specific MEI agents may serve as health companions, reminding individuals about medication, tracking vital signs, and offering empathetic engagement. These applications require robust data governance and ethical oversight to ensure privacy, accuracy, and trust. As digital twin frameworks evolve, MEI can bridge the gap between clinical expertise and patient-centered care, fostering resilience and continuity in health systems.

### 4.5 Governance and Leadership

MEI embodiments can enhance governance by enabling leaders to engage across multiple forums, synthesize policy data, and interact with constituents in real time. AI-enabled agents may attend summits, participate in deliberations, and relay feedback to the original individual. Acosta & Hughes (2025) propose an adaptive governance model where AI supports anticipatory decision-making and inclusive consultation.

Boch and Thomas (2025) argues that leadership in the age of AI requires a redefinition of authority, trust, and ethical stewardship. MEI agents, when aligned with leadership values and equipped with escalation protocols, can act as strategic partners rather than mere tools. This distributed leadership model promotes transparency, responsiveness, and symbolic presence across diverse governance contexts.

### 4.6 Media and Entertainment

In media and entertainment, MEI agents can participate in interviews, performances, and public engagements, offering audiences access to a replicated personality. The World Economic Forum (2025) highlights generative AI's role in augmenting creativity, optimizing content distribution, and enhancing audience engagement. MEI embodiments may serve as virtual performers, collaborators, or commentators, expanding the reach of public figures.

Sengar et al., (2025) notes that generative AI can reduce production costs and democratize content creation, enabling high-quality output from diverse creators. MEI agents, when ethically endorsed and contextually adapted, can redefine celebrity presence and audience interaction. These applications must navigate intellectual property rights, authenticity concerns, and commercialization boundaries to ensure responsible innovation.

## 5. Challenges and Ethical Considerations

While the potential of MEI is profound, its deployment raises critical challenges that must be addressed to ensure social legitimacy and ethical integrity.

### 5.1 Authenticity and Identity

The authenticity of MEI embodiments raises fundamental ontological questions. Are these entities extensions of selfhood or autonomous simulations? From a Lockean continuity perspective (Locke, 1690/1975), authenticity depends on whether MEI shares memory and decision continuity with the original. From a phenomenological perspective (Zahavi, 2005), authenticity is relational—determined by how others experience the interaction. MEI thus destabilizes traditional boundaries of presence by creating entities that are both "the self" and "not the self." This duality requires a rethinking of what constitutes authentic interaction in digitally mediated societies.

### 5.2 Control and Consent

MEI systems raise critical concerns about control and autonomy. Who defines the boundaries of an MEI's decision-making authority, and what happens when its actions diverge from the individual's intent? These questions underscore the need for robust consent frameworks—not only at the point of deployment but throughout the MEI's operational lifecycle. Without such safeguards, accountability becomes ambiguous, and the risk of ethical and legal violations increases.

Control mechanisms must balance autonomy with accountability. A deontological lens (Kant, 1785/1993) stresses non-negotiable principles: individuals must retain ultimate veto power over their MEI embodiments. A utilitarian perspective (Mill, 1863/2001), by contrast, emphasizes maximizing efficiency and benefit—even if agents operate semi-independently. Consent protocols must therefore extend beyond initial deployment to include ongoing, revocable consent governing how MEI agents act, adapt, and are socially perceived.

### 5.3 Legal and Responsibility Issues

The actions of MEI embodiments may carry legal or contractual consequences. Determining liability—whether it rests with the original individual, the AI system, or the deploying organization—poses significant legal challenges. Jurisdictions will need to adapt laws around agency, responsibility, and personhood.

Current legal frameworks recognize natural persons and, in limited cases, corporate entities as bearers of rights and liabilities. MEI challenges this by distributing agency across multiple embodiments. If a robotic copy signs a contract or causes harm, liability assignment becomes complex. Jurisdictions may need to adopt models similar to those used in corporate liability or autonomous vehicle regulation (Calo, 2016), where accountability is distributed but anchored to a human source.

**5.4 Privacy and Data Security**

Because MEI relies on comprehensive datasets of personal characteristics, behaviors, and preferences, ensuring data privacy is paramount. Unauthorized access or manipulation of MEI systems could result in identity theft at a far deeper level than traditional data breaches.

Because MEI systems require highly intimate datasets (language, gestures, preferences, emotional tendencies), their misuse could lead to profound forms of identity theft. Under frameworks like the General Data Protection Regulation (GDPR) (European Union, 2016), MEI would involve sensitive "biometric and behavioral data," requiring strict consent, data minimization, and right-to-erasure protocols. Without safeguards, MEI could be weaponized for manipulation, surveillance, or fraudulent activity.

**5.5 Psychological and Social Impact**

The use of MEI is anticipated to influence interpersonal relationships, potentially diminishing the value of authentic physical presence. Families, for example, may experience confusion between genuine and mediated interactions. On a societal level, widespread adoption could reshape norms of intimacy, authenticity, and trust.

On an interpersonal level, MEI may blur distinctions between genuine and mediated presence, altering family dynamics, romantic relationships, and professional trust. Socially, widespread MEI adoption could normalize interactions with copies, potentially reducing the value of unmediated human presence. Critical here is whether MEI promotes resilience and inclusion, or whether it exacerbates alienation and dependency on artificial proxies.

Collectively, these ethical, legal, and psychological considerations suggest that MEI cannot be treated as a purely technical innovation. It is a sociotechnical system that demands governance frameworks combining ethical philosophy, data protection law, and applied AI regulation. Without such safeguards, the risks of fragmentation, exploitation, and inequity may outweigh the potential benefits.

**6. Discussion**

The notion of *Multi-Existence Identity (MEI)* extends contemporary debates on digital embodiment, identity replication, and the augmentation of human presence. Existing technologies such as digital twins, telepresence systems, and conversational AI provide important stepping stones, but they remain limited to narrow functional domains. MEI advances this discourse by envisioning a holistic integration of mental coherence, affective coherence, and embodied interaction.

While MEI presents a visionary framework, its realization depends on the convergence of several emerging technologies. Current AI systems such as large language models (e.g., GPT-4) demonstrate impressive linguistic and behavioral mimicry, enabling personality modeling through fine-tuned datasets. However, the long-term sustainability of such systems remains uncertain and warrants further investigation. Cognitive simulation is partially feasible via reinforcement learning and affective computing, though limitations remain in modeling non-verbal cues, emotional nuance, and long-term memory continuity. Embodiment channels—digital avatars and robotic interfaces—are advancing rapidly, with humanoid robots capable of basic social interaction and avatars integrated into metaverse platforms. However, real-time synchronization across multiple embodiments remains a major technical challenge. Achieving coherence and continuity requires robust cloud-based architectures, low-latency communication, and adaptive control systems that can reconcile divergent experiences into a unified memory stream. In the near term, MEI prototypes may be viable in narrow domains, such as email response agents, virtual meeting avatars, or robotic companions for scripted interactions. These limited applications could serve as testbeds for refining thought-style alignment and synchronization mechanisms before expanding into more complex, socially embedded embodiments.

From a sociotechnical perspective, MEI illustrates the convergence of distributed cognition (Hutchins, 1995) and embodied artificial intelligence (Dautenhahn, 2007), thereby reframing the human as a *distributed entity* whose presence can be multiplied across physical and digital spaces. This raises both opportunities and risks. On the one hand, MEI may contribute to expanded productivity, social continuity, and accessibility to knowledge and resources. On the other hand, it may destabilize traditional understandings of personhood, responsibility, and authentic interaction.

From a cultural and artistic perspective, MEI can also be framed through metaphors of performance and rhythm. Like the Bunraku puppet theatre, where the puppeteer's presence is rhythmically enacted yet socially concealed, MEI agents may render identity simultaneously visible and invisible. Similarly, improvisational arts highlight how tacit rhythms guide interaction—an analogy that illuminates MEI's potential to sustain identity continuity not by rigid control but by flexible, culturally embedded enactment.

### 6.1 Empirical Roadmap for Operationalizing MEI

To empirically evaluate MEI, a phased roadmap is proposed that builds on precedents from avatar studies, social robotics, and distributed cognition research. Although the article is primarily conceptual, each phase outlines testable prototypes and specifies datasets, evaluation criteria, and ethical safeguards. Together, the phases move MEI from speculative vision to a structured research program.

- A. Phase 1 — Personality Model Development
    Using natural language processing (NLP) tools such as GPT fine-tuning and BERT embeddings, alongside behavioral data (emails, chat logs, decision logs, affective responses),

an individual-specific model is trained. This phase draws on personality computing (Kosinski et al., 2013) and conversational agent fine-tuning (Zhou et al., 2020).

- Data: Textual corpora from 20–30 volunteers, ~50,000 words each; anonymised and collected with consent.

- Evaluation: Alignment with Big Five Inventory (BFI-2); linguistic mimicry accuracy via embedding similarity (e.g., SBERT); emotional response fidelity.

- Baseline: A generic fine-tuned GPT-style assistant without personalization.

- Analysis: Paired t-tests and effect sizes comparing model vs. baseline.

B. Phase 2 — Digital Avatar Deployment

The trained personality model is instantiated in controlled virtual environments (Unity, Unreal, or equivalent). Research on telepresence avatars (Wang, 2018; Oh et al., 2018) demonstrates that acceptance and authenticity can be measured through behavioral coding and self-report.

- Contexts: Virtual meetings, metaverse platforms, online classrooms.

- Evaluation: User trust and perceived social presence (survey scales); behavioral congruence with the original individual; interaction fluidity and latency.

- Baseline: Conventional multi-device assistants (e.g., Alexa, Siri).

- Analysis: Repeated-measures ANOVA comparing MEI-driven avatars with baselines.

C. Phase 3 — Synchronization Mechanisms

MEI agents and the original individual are placed in parallel scenarios, testing the capacity to maintain coherence across distributed embodiments. This builds on distributed cognition (Hutchins, 1995) and multi-agent reinforcement learning.

- Tools: Decision trace analysis, memory stream integration, emotional state alignment.

- Evaluation: Decision coherence across embodiments; synchronization latency (ms); override success rate.

- Analysis: Longitudinal tracking and statistical comparison of divergence over time.

D. Phase 4 — Robotic Embodiment Trials

Selected personality models are embedded into humanoid or service robots (e.g., NAO, Pepper), drawing on field studies in social robotics (Breazeal, 2003; Dautenhahn, 2007). Deployments occur in low-risk environments such as companionship, customer service, or education.

- Method: Ethnographic observation, interaction coding, participant interviews.
- Evaluation: Observed behavioral fidelity; social acceptance and comfort levels; ethical concerns raised by participants.
- Analysis: Independent-samples t-tests for trust and comfort scores; chi-square for task success frequencies.

E. Phase 5 — Ethical and Legal Stress-Testing

Inspired by autonomous vehicle trials and AI governance frameworks (Floridi et al., 2018), this phase subjects MEI to ethically sensitive and legally complex scenarios.

- Scenarios: Contract signing, caregiving decisions, public representation.
- Procedures: Delphi panels with ethicists, legal scholars, and technologists; GDPR-aligned audits; consent-protocol stress tests.
- Evaluation: Legal feasibility index; ethical acceptability scores; robustness of consent and override mechanisms.
- Analysis: Expert ratings with inter-rater reliability (Cohen's κ); descriptive statistics on compliance.

F. Ethics and Governance Safeguards

Informed consent is obtained at every stage; participants retain ownership of their raw data. Only minimal anonymised subsets are stored, encrypted locally, with privacy-enhancing techniques (e.g., differential privacy) applied where possible. Institutional ethics approval governs all studies, and participants may withdraw at any point with immediate data deletion.

## 7. Ethics and Data Governance for MEI

Because MEI relies on personal behavioral traces, ethical safeguards must be treated as integral to system design rather than as afterthoughts. This section specifies concrete governance mechanisms across the data lifecycle, combining technical and procedural safeguards.

- **Consent and Participation Model**

Participants provide *tiered consent* (Kaye et al., 2015), explicitly choosing which data types (e.g., text logs, audio samples, decision records) may be used. Consent is revocable at any time: withdrawal triggers immediate deletion of all linked datasets. A dynamic consent dashboard allows participants to monitor, pause, or revoke data contributions (Kaye et al., 2015).

- **Data Minimization and Purpose Limitation**

Only the minimal subset of data necessary for training a personality model is collected. For instance, 5–10 representative conversations may suffice to seed an initial model rather than archiving entire communication histories. Data are processed for clearly defined research purposes (model training, validation, evaluation) and not repurposed without renewed consent (Floridi et al., 2018).

- **Storage and Access Controls**

All raw data are anonymized and stored in encrypted local servers under institutional governance; cloud storage is avoided unless encrypted end-to-end. Access is restricted to authorized researchers, logged, and audited regularly. Derived personality embeddings (vector representations) are separated from raw text and stored without direct identifiers (Floridi et al., 2018).

- **Privacy-Preserving Model Training**

*Federated learning* will be explored to allow models to train locally on participant devices, transmitting only model updates rather than raw data (McMahan et al., 2017). *Differential privacy* mechanisms (e.g., noise injection in gradient updates) are applied to reduce the risk of re-identification (Dwork & Roth, 2014). All experiments comply with relevant data protection regulations (e.g., GDPR, New Zealand Privacy Act).

- **Ethical Review and Oversight**

Each phase of MEI experimentation is reviewed by an accredited Institutional Review Board (IRB) or equivalent ethics committee, following the principles of the Belmont Report (1979). Periodic audits by independent ethicists and legal scholars ensure adherence to ethical standards beyond compliance. A "red-team" process subjects MEI to adversarial scenarios to test resilience against misuse (e.g., impersonation, data leakage).

- **Participant Rights and Transparency**

Participants retain the right to access, correct, or delete their data and to receive a plain-language summary of how their data informed the MEI model. Results of empirical phases are communicated back to participants to maintain transparency and reciprocal accountability (Floridi et al., 2018).

By embedding these governance safeguards into the empirical roadmap, MEI advances not only technical feasibility but also social legitimacy. Privacy-preserving computation and continuous ethical oversight ensure that identity replication does not compromise the dignity or autonomy of participants.

# 8. Conclusion

This article has proposed the conceptual framework of MEI, whereby an individual's cognitive, behavioral, and personality attributes are replicated into AI-driven embodiments capable of acting across multiple contexts simultaneously. Building on prior work in digital twins and telepresence, MEI extends these traditions into a unified framework and expands them into a more ambitious paradigm: distributed human identity. We propose MEI as a next-generation framework that redefines how identity and agency are distributed across digital and physical realms.

Applications of MEI span professional, familial, educational, healthcare, and governance contexts, offering unprecedented possibilities for extending human presence and productivity. However, these opportunities are inseparable from challenges surrounding authenticity, control, legal accountability, privacy, and the psychological meaning of presence.

The contribution of this article lies in presenting MEI as a speculative yet plausible trajectory for human-technology integration. By outlining a conceptual framework and highlighting both potential applications and ethical risks, it provides a foundation for future scholarly debate and empirical inquiry. Ultimately, the realization of MEI demands not only technological innovation but also careful negotiation of the values and norms that define what it means to be human in an age of distributed existence.

In the long term, MEI suggests the possibility of reconfiguring the boundaries of human presence, enabling individuals to participate in multiple spheres of life simultaneously. This has the potential to contribute to advancements in productivity, caregiving, education, and governance. However, it also risks fragmenting identity, diluting interpersonal authenticity, and creating new forms of digital inequality. If MEI becomes a privilege of the technologically empowered, it carries the risk of contributing to increased social divides. Furthermore, the psychological impact of interacting with replicated selves—both for the individual and others—requires careful study. As MEI agents become more autonomous, society must grapple with questions of moral agency, legal responsibility, and the sanctity of human uniqueness. More broadly, MEI challenges the pursuit of certainty and algorithmic control that often dominates AI discourse. Rather than presuming that identity can be exhaustively modeled or governed, MEI foregrounds dialogic, situated, and tacit forms of presence that resist reduction to purely computational logic. This opens pathways for rethinking human–AI relations in less deterministic and more relationally grounded terms. In resisting the pursuit of certainty and algorithmic control, MEI underscores that identity cannot be exhaustively codified. Instead, it foregrounds dialogic and tacit modes of presence that are enacted rather than computed, positioning MEI as a corrective to reductionist models of human–AI relations. These risks underscore the need for inclusive, ethical, and interdisciplinary governance frameworks to guide MEI's development.

**Glossary of Key Terms**

- Identity Fidelity – The degree to which MEI agents replicate the original individual's cognitive style, emotional patterns, and behavioral traits (includes cognitive fidelity and behavioral continuity).
- Tacit Engagement – The embodied, relational, and culturally situated dimensions of presence that extend beyond explicit data or representation.
- Affective Resonance – The ability of MEI agents to reproduce emotional rhythms, tones, and relational cues that align with the original person.
- Embodiment Channels – The three operational forms of MEI: digital avatars, robotic embodiments, and agentic software agents.
- Agentic Autonomy – The bounded decision-making capacity of MEI agents, guided by personality parameters and synchronization protocols.
- Distributed Cognition – A theory of cognition as extended across people, tools, and environments, foundational to MEI's design.
- Relational Authenticity – The extent to which interactions with MEI agents feel genuine and socially faithful to the original individual.
- Identity Synchronization – The system ensuring that MEI agents remain aligned with the original person's intentions, memories, and values.
- Replicated Identity – An MEI embodiment designed to act as the individual by mirroring cognitive and emotional style, distinct from generic simulated agents.
- Embodied AI – AI instantiated in physical or virtual forms that interact with humans through naturalistic and socially meaningful engagement.

**Appendix A: MEI Agent Instantiation Architecture**

The instantiation of MEI agents follows a modular architecture designed to replicate cognitive, emotional, and behavioral attributes of an individual across distributed embodiments. The architecture integrates personality modeling, cognitive simulation, embodiment channels, and synchronization protocols.

**A.1 System Architecture Overview**

**Input Layer**

- Data sources: linguistic corpora (emails, chat logs), behavioral logs, affective responses
- Tools: NLP pipelines (e.g., BERT, GPT), psychometric profiling (e.g., Big Five Inventory)

→ **Personality Modeling Module**

- Constructs a dynamic representation of the individual's cognitive style, emotional rhythms, and decision heuristics

- Output: a personalized agent model with embedded values and interaction patterns

→ **Embodiment Channels**

- Digital Avatars: deployed in virtual platforms (e.g., meetings, metaverse)

- Robotic Embodiments: instantiated in physical robots (e.g., NAO, Pepper)

- Agentic Software Agents: autonomous programs for task execution (e.g., negotiation, scheduling)

→ **Synchronization Layer**

- Maintains coherence across embodiments via bi-directional memory integration and override protocols

- Ensures alignment of decisions, emotional states, and contextual awareness

### A.2 Deployment Logic

A. Task Selection: The individual defines which tasks require direct presence and which can be delegated.

B. Agent Activation: MEI agents are instantiated in selected channels with bounded autonomy.

C. Feedback Loop: Agents relay summaries and decisions to the original individual.

D. Override & Escalation: Critical decisions trigger real-time consultation or override mechanisms.

E. Memory Integration: Experiences from all embodiments are synchronized into a unified memory stream.

### A.3 Technical Stack (Illustrative)

**Table 4:** Core Components and Technologies for MEI Agent Instantiation

| Component | Technologies / Methods |
|---|---|
| NLP and Personality Model | GPT fine-tuning, BERT embeddings, psychometrics |
| Embodiment Interface | Unity, Unreal Engine, ROS, humanoid robotics |

| Component | Technologies / Methods |
|---|---|
| Synchronization Layer | Multi-agent RL, distributed cognition protocols |
| Ethical Safeguards | GDPR compliance, consent protocols, audit logs |

This appendix provides a foundational blueprint for MEI implementation and can be expanded in future empirical studies or prototype development.